\documentclass[11pt,fleqn]{article}
\usepackage{epstopdf} 
\usepackage{graphicx}
\usepackage[a4paper,left=2cm,right=2cm]{geometry}
\usepackage{amsmath}
\usepackage{float}
\usepackage[latin1]{inputenc}
\usepackage{amsmath}
\usepackage{amsfonts}
\usepackage{amssymb}
\usepackage{float}
\usepackage[toc,page]{appendix}
\floatstyle{boxed} 
\restylefloat{figure}

\newcommand{\be}{\begin{equation}}
\newcommand{\ee}{\end{equation}}
\newcommand{\bea}{\begin{eqnarray}}
\newcommand{\eea}{\end{eqnarray}}

\begin{document}
\title{Search for an exotic-leptonic solution to observed anomalies in lepton universality observables in B decays and more.\footnote{Preprint No. : HRI-RECAPP-2018-010}}
\author{Lobsang Dhargyal \\\\\ Regional Centre for Accelerator-based Particle Physics,\\\\\ Harish-Chandra Research Institute, HBNI, Jhusi, Allahabad - 211019, India}
\date{July 31, 2018.}


\maketitle
\begin{abstract}

In this work we extend the SM by introducing only exotic scalars and leptons and show that within the reasonable error limits, both the observed anomalies in $R(D^{(*)})$ and $R_{K^{(*)}}$ can be explained along with satisfying all the constrains from nuetral meson oscillations, precision Z-pole data etc. In a trivial extension of our model with addition of three heavy right handed neutrinos, explaining the small masses of neutrinos and generation of Baryon excess via leptogenesis is possible as well. The disagreement between SM prediction and experimental data in muon (g-2) measurements can be reduced from 3.6$\sigma$ to around 2$\sigma$ in this model. Also our model has enough new particles for the scalar singlet DM to interact and generate enough annihilation to avoid over abundance problem (exotic portal), unlike the SM Higgs portal which has been ruled out.

\end{abstract}
\maketitle

\section{Introduction.}

Even though LHC reporting observation of no new particles beyond the standard-model (SM) Higgs in direct searches, some observables in flavor sector show tension with SM prediction up to about 4$\sigma$ in some cases. It may be an indication that new-physics (NP) scale is close to the SM scale and so a precision machine is better equipped than an energy frontier machine to probe the nature of NP that lies beyond present SM. Besides the well known short comings of SM such as it is unable to explain the existence of small but non-zero neutrino masses, dark-energy (DE), dark-matter (DM) and observed Baryon excess in the universe, some experiments in flavor sector have reported intriguing deviations in lepton universality observables such as $R(D^{(*)}) = \frac{Br(B \rightarrow D^{(*)} \tau \nu_{\tau})}{Br(B \rightarrow D^{(*)} l \nu_{l})}$ \cite{our3-ref6}\cite{our3-ref17}\cite{our3-ref18}\cite{our3-ref19}\cite{our3-ref20}\cite{our3-ref21}, $R_{K^{(*)}} = \frac{Br(B \rightarrow K^{(*)}\mu^{+}\mu^{-})}{Br(B \rightarrow K^{(*)}e^{+}e^{-})}$ \cite{our2-ref4} and muon (g-2) \cite{our2-ref1}. For the anomalies in $R_{K^{(*)}}$ and related observables, in \cite{our2-ref9} it has been determined that the combined global data, which is about 4$\sigma$ deviated from SM prediction \cite{our2-ref9}, is best fitted by a NP with $-0.81 \leq C^{NP}_{9} = -C^{NP}_{10} \leq -0.51$ at 1$\sigma$. For the $R(D^{(*)})$ observables the HFAG global average is given as \cite{our3-ref22}
\be
R(D)^{Exp} = 0.407(0.039)(0.024)
\ee
and
\be
R(D^{*})^{Exp} = 0.304(0.013)(0.007),
\ee
amounting to about 4.1$\sigma$ deviation from SM prediction and it has been shown that one of the best fit NP model would be one that add coherently to the SM effective four current in this observables \cite{our3-ref23}. Then there is the reported deviation from SM prediction in muon (g-2) with present global average of the deviation given as
\be
\delta a_{\mu} = a^{Exp.}_{\mu} - a^{NP}_{\mu} = 288(69)(49)\times 10^{-11},
\ee
which is about 3.6$\sigma$ away from SM prediction \cite{our2-ref1}. In this work we will propose a NP model that will be able to generate $C^{NP}_{9} = -C^{NP}_{10}$ via box-loop to explain the anomalies in $R_{K^{(*)}}$ and related observables and a NP Wilson coefficient that add coherently with the SM effective four current, also via box-loop, to explain the $R(D^{(*)})$ anomalies as well as explaining the anomaly in muon (g-2) within 2$\sigma$, smallness of neutrino masses and baryon-genesis. This work is organized as follows, in section \ref{mod-det} the details of the NP model is given, in section \ref{sect:Impl-const} the implications of the NP model to flavor physics observables along with constrains on NP parameters from flavor precision data. And in section \ref{sect:conclusions} we conclude.

\section{Model details.}
\label{mod-det} 

In some recent works it has been shown that NP models with exotic leptons and scalars contributing to $R(D^{(*)})$ \cite{our3}, $R_{K^{(*)}}$ \cite{our2}(at box-loop level) and muon (g-2) \cite{our1} will be able to resolve the reported anomalies in those observables within the error limits. In this work we would like to propose an extension of SM which will be able to resolve the anomalies in both $R(D^{(*)})$ and $R_{K^{(*)}}$ as well as smallness of nutrino masses and Baryon-genesis. We add to SM two $SU(2)_{L}$ doublets $L_{1 L} = (N_{1 L},E_{1 L})$ and $L_{2 R} = (N_{2 R},E_{2 R})$ along with $SU(2)_{L}$ singlets $E_{1 R}$ and $E_{2 L}$, they are all leptons carrying same $U(1)_{Y}$ charges as the SM lepton doublets and singlets respectively. We also add three SM singlet right handed neutrinos $(N_{eR}, N_{\mu R}, N_{\tau R})$ to the SM lepton content to generate small neutrino masses at loop level. Where the subscripts e, $\mu$ and $\tau$ denotes the SM lepton number carried by the heavy right handed neutrinos. Since new leptons form a vector like pairs under the relevant SM gauge groups, the model is free of axial anomaly. The collider signatures of a locally gauged lepton number extension of SM with similar new particle content as our model is proposed in \cite{WFChang1}\cite{WFChang2}, but here we will keep the lepton number to be a global gauge as in the SM case. All the new leptons are assumed to be odd under the $\mathcal{Z}_{2}$. Also to SM Higgs, we add two scalar leptoquarks one $SU(2)_{L}$ singlet $\phi_{LQ}$ and one $SU(2)_{L}$ doublet $\eta_{LQ}$ along with an inert-doublet $\eta$ and a singlet S, with all the new scalars also being odd under the $\mathcal{Z}_{2}$. One more real scalar singlet under the SM gauge group $\phi$ is also added, which is even under the $\mathcal{Z}_{2}$ and which develop a non-zero VEV to give masses to the new leptons. In Table \ref{tab1} and Table \ref{tab2} we have shown the charge assignments of the new leptons and new scalars respectively.
\begin{table}[h!]
\begin{center}
\begin{tabular}[b]{|c|c|c|c|c|} \hline
Particles & $SU(3)_{c}$ & $SU(2)_{L}$ & $U(1)_{Y}$ & $\mathcal{Z}_{2}$ \\
\hline\hline
$L_{1 L}$ & 1 & 2 & -1/2  & -1 \\
\hline
$E_{1 R}$ & 1 & 1 & -1  & -1 \\
\hline
$E_{2 L}$ & 1 & 1 & -1  & -1 \\
\hline
$L_{2 R}$ & 1 & 2 & -1/2  & -1 \\
\hline
$N_{i R}$ & 1 & 1 & 0  & -1 \\
\hline
\end{tabular}
\end{center}
\caption{The charge assignments of new leptons under the SM gauge group and $\mathcal{Z}_{2}$ and $i = e, \mu, \tau$.}
\label{tab1}
\end{table}

\subsection{Yukawa Interactions.}
\label{sub:Yukawa}

The most general Yukawa interaction terms that are invariant under the full symmetries of the model can be written down as
\be
\begin{split}
\mathcal{L}_{Yukawa} = Y_{2}\bar{L}_{1 L}H E_{1 R} + Y_{3}\bar{L}_{2 R}H E_{2 L} + \lambda_{3}\bar{L}_{1 L}L_{2 R}\phi + \lambda_{4} \bar{E}_{2 L}E_{1 R}\phi \\
+ \sum_{i=u}^{i=t}h_{2i}\bar{Q}_{iL}L_{2 R}\phi_{LQ} + \sum_{j=e}^{j=\tau}h_{2j}\bar{L}_{jL}L_{2 R}S + \sum_{i=u}^{i=t}h_{1i}\bar{Q}_{iL}\eta_{LQ}E_{1 R} + \sum_{j=e}^{j=\tau}h_{1j}\bar{L}_{jL}\eta E_{1 R}.
\end{split}
\label{Yuk-Eq}
\ee
After SM Higgs H and the new scalar $\phi$ develops a non-zero VEV $v_{0}$ and $v_{1}$ respectively, we have the mass matrix of the new charged leptons given as
\be
M_{E} = \frac{1}{\sqrt{2}}\begin{bmatrix} Y_{2}v_{0} & \lambda_{4}v_{1} \\ \lambda_{3}v_{1} & Y_{3}v_{0} \end{bmatrix}
\ee
and matrix of new non-neutrino neutral leptons given as
\be
M_{N} = \frac{1}{\sqrt{2}}\begin{bmatrix} 0 & \lambda_{3}v_{1} \\ \lambda_{3}v_{1} & 0 \end{bmatrix}.
\ee
In this work we take the limit $Y_{2}v_{0} = Y_{3}v_{0} \approx m_{l} << \lambda_{3}v_{1} = \lambda_{4}v_{1} \approx m_{E}$ where $m_{l}$ is mass in the order of SM charged lepton masses. Then as shown in \cite{WFChang1}, both the matrices of new charged leptons and new nuetral leptons are diagonalized by same rotation matrix
\be
U = \frac{1}{\sqrt{2}} \begin{bmatrix} 1 & 1 \\ -1 & 1 \end{bmatrix}
\ee
with degenerate masses for the nuetral leptons at tree level and mass difference of order $2m_{l}$ for the new charged leptons, so in the relevant scenario where we take the new lepton masses well above the scale of SM lepton masses, we can take $m_{N} \approx m_{E_{h}} \approx m_{E_{l}}$, i.e in the limit stipulated above we can take the neutral lepton and charged leptons having nearly degenerate masses, where subscript h and l denote heavy and light particle respectively.
\begin{table}[h!]
\begin{center}
\begin{tabular}[b]{|c|c|c|c|c|} \hline
Particles & $SU(3)_{c}$ & $SU(2)_{L}$ & $U(1)_{Y}$ & $\mathcal{Z}_{2}$ \\
\hline\hline
$\phi_{LQ}$ & 3 & 1 & +2/3 & -1 \\
\hline
$\eta_{LQ}$ & 3 & 2 & 7/6 & -1 \\
\hline
$\eta$ & 1 & 2 & 1/2 & -1 \\
\hline
S & 1 & 1 & 0 & -1 \\
\hline
$\phi$ & 1 & 1 & 0 & +1 \\
\hline
\end{tabular}
\end{center}
\caption{The charge assignments of new scalars under the SM gauge group and $\mathcal{Z}_{2}$.}
\label{tab2}
\end{table}
SM gauge interactions and collider productions and decay signatures of our model is same as those given in \cite{WFChang1}. In the near degenerate masses for the new exotic neutral and charged leptons as well as near degenerate masses for the charged Higgs and heavier nuetral Higgs of inert-doublet cases, as will be assumed in this work, the contributions to S and T parameters are $\Delta T^{NP} \approx 0$ and $\Delta S^{NP} \approx 0.106$ which is well within the present experimental limit of $\Delta T^{Exp.} < 0.27$ and $\Delta S^{Exp.} < 0.22$. Also as shown in \cite{WFChang1}, in our case where the SM Higgs coupling to the new exotic leptons are at same order as the SM lepton couplings to the SM Higgs, the contributions to the new exotic leptons and charged Higgs to $h \rightarrow \gamma \gamma$ is within the experimental limit even for the light charged Higgs to have Yukawa couplings of $\mathcal{O}(1)$.

\section{Constrains and implications in flavor physics.}
\label{sect:Impl-const}

In SM if we multiply only the first two rows of the CKM matrix elements with -1, there is no observable that can detect this sign change. But here in our model, as will be shown in the following paragraphs, this change in relative sign between rows of CKM matrix elements have observable effect. Here we will fix the angles of CKM matrix elements as $\pi \leq \theta_{12} \leq \frac{3\pi}{2}$ and $\frac{3\pi}{2} \leq \theta_{13}, \theta_{23} \leq 2\pi$, i.e the signs of the first two rows of CKM matrix elements are changed relative to the third row compared to the usual convention where all the angles are fixed in the first quadrant \cite{our3}. The Yukawa couplings in the exotic lepton sector taken as $h_{2e, 2\mu} << h_{2\tau} \approx 2\sqrt{\pi}$ and $h_{1e, 1\tau} << h_{1\mu} \approx 2\sqrt{\pi}$ is favored by the reported anomalies. The Yukawa couplings in the down quark sector in mass diagonal states can be expressed as
\be
h_{(1,2)i}^{'} = \sum^{j=t}_{j=u}h_{(1,2)j}V_{ji}
\ee
where $V_{ij}$ are CKM matrix elements and $i = d,\ s,\ b$\footnote{in this notation we take $h_{u, c, t} = h_{d, s, b}$ respectively}. Since $K^{0}-\bar{K}^{0}$ and $B^{0}-\bar{B}^{0}$ are very precisely measured and there being no deviations observed in this modes, these data can be accommodated easily if we impose
\be
h^{'}_{(1,2)d} = \sum^{j=t}_{j=u}h_{(1,2)j}V_{jd} = 0.
\label{hd}
\ee
In this work the Yukawa couplings are assumed to satisfy the above conditions. It can be shown that the constrains from $B^{0}_{s}-\bar{B}^{0}_{s}$ oscillation and $R(D^{(*)})$ data along with the condition $h^{'}_{2d} = -V_{ud}h_{2d} - V_{cd}h_{2s} + V_{td}h_{2b} = 0$(where sign change of the first two rows of CKM elements are shown explicitely) can be satisfied for $h_{2b} = 3.52 < 2\sqrt{\pi}$ and $h_{2s} = -\frac{h_{2b}}{21.588}$ with $Re(h_{2d}) = -8.402\times 10^{-3}$ and $Im(h_{2d}) = -0.0119$ \cite{our3}.\\
Similarly $h^{'}_{1d} = -V_{ud}h_{1d} - V_{cd}h_{1s} + V_{td}h_{1b} = 0$ can be satisfied along with explaining the $R_{K^{(*)}}$ data for $h_{1s} = h_{1b} = \frac{2\sqrt{\pi}}{21.588}$ with $Re(h_{1d}) = 0.039$ and $Im(h_{1d}) = -5.51\times 10^{-4}$. In the above calculations we have taken the values of  the Wolfenstien parameters of CKM from PDG \cite{our2-ref1}.

\subsection{Neutral meson oscillation.}
\label{subsect:Osc}

Like in the SM, both $E_{1R}$ and $L_{2R}$ can contribute to the $B^{0}_{s}-\bar{B}_{s}^{0}$ oscillation at box loop level. Their contributions can be expressed as \cite{our2}\cite{our3}
\be
\Delta M_{B^{0}_{s}}^{NP} = \frac{1}{3}m_{B_{s}^{0}}f^{2}_{B^{0}_{s}}B(\mu)\times (C^{E_{1R}}_{B_{s}^{0}} + C^{L_{2R}}_{B_{s}^{0}})
\ee
where $m_{B_{s}^{0}}$ is the measured mass of the $B_{s}^{0}$ with $f^{2}_{B^{0}_{s}}$ and $B(\mu)$ are the decay constant and QCD scale correction factor respectively, their values are taken from \cite{our3-ref12}\cite{our3-ref14}. The $C^{E_{1R}}_{B_{s}^{0}}$ and $C^{L_{2R}}_{B_{s}^{0}}$ can be expressed as
\be
C^{E_{1R}}_{B_{s}^{0}}(C^{L_{2R}}_{B_{s}^{0}}) = \frac{(h^{'}_{1s(2s)}h^{'}_{1b(2b)})^{2}}{128\pi^{2}m^{2}_{E}}S(x,x)
\ee
where $S(x,x)$ being Inami-Lim functions, see \ref{subsect:bctnu} for detail, and $x = \frac{m_{LQ}^{2}}{m_{E}^{2}}$ with $m_{LQ}$ denoting the mass of the leptoquark involved.
\begin{figure}[h!]
\begin{minipage}[t]{0.48\textwidth}
\hspace{0.4cm}
\includegraphics[width=2\linewidth, height=8cm]{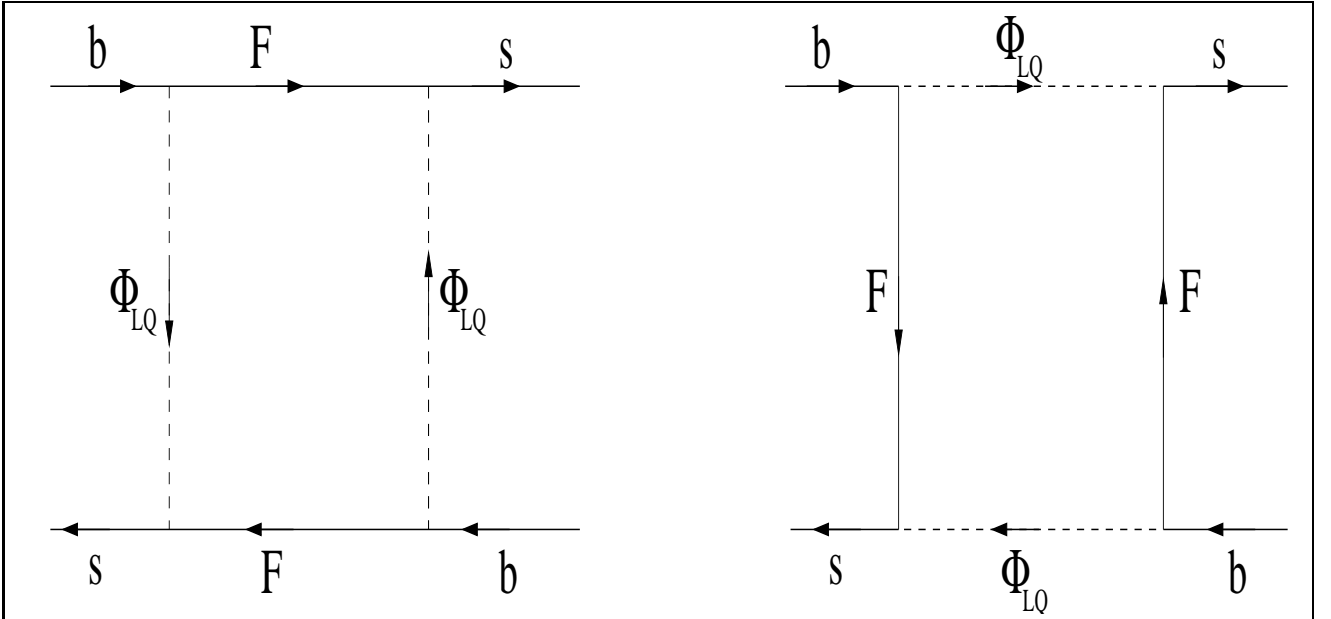}
\end{minipage}
\caption{Contributions to the $B^{0}_{s}-\bar{B}_{s}^{0}$ oscillation from the new particles at box loop level, where $F$ denoting the relevant new fermions involved.}
\label{Fig1:fig2}
\end{figure}
With benchmark values of masses taken as $m_{LQ} = 900$ GeV and $m_{N} = m_{E} = 300$ GeV, we get $Re(\Delta M_{B^{0}_{s}}^{NP}) = 1.337 ps^{-1}$ which is within the 1.1$\sigma$ of the error in the latest SM prediction given as $M_{B^{0}_{s}}^{SM} = (20.01 \pm 1.25) ps^{-1}$. Due to $(h^{'*}_{1s(2s)}h^{'}_{1b(2b)})^{2}$ being complex, there is also an imaginary component of $\Delta M_{B^{0}_{s}}^{NP}$ which can contribute to CP violation observables in the $B^{0}_{s}-\bar{B}_{s}^{0}$ oscillation. It turn out that with values of the Yukawa couplings given in the section \ref{sect:Impl-const}, $Im[(h^{'*}_{1s}h^{'}_{1b})^{2}] \approx \mathcal{O}(10^{-6})$ and $Im[(h^{'*}_{2s}h^{'}_{2b})^{2}] \approx -1.08\times 10^{-3}$, so contribution due to $E_{1R}$ is negligible compared to $L_{2R}$. With the benchmark values for the parameters, we get $Im(\Delta M_{B^{0}_{s}}^{NP}) \approx -0.715\times\Gamma^{Exp}_{B_{s}^{0}}$ which gives $\frac{RE(\epsilon^{NP})}{1+|\epsilon^{NP}|^{2}} \approx \frac{-Im(\Delta M_{B_{s}^{0}}^{NP})\times \Gamma^{Exp.}_{B_{s}^{0}}}{4(\Delta M_{B^{0}_{s}}^{Exp.})^{2}} \approx 1.050\times 10^{-5}$ compared to $\frac{RE(\epsilon^{Exp.})}{1+|\epsilon^{Exp.}|^{2}} \approx (-1.5 \pm 7)\times 10^{-4}$, the NP contributions is an order of magnitude smaller than the present experimental limit. For the $D^{0}-\bar{D}^{0}$ oscillation, we have at 2$\sigma$ experimental bound as $|C^{Exp.}_{D^{0}}| < 2.07\times 10^{-7}\ TeV^{-2}$ compared to the NP contribution given as $|C^{NP}_{D^{0}}| = 6.393\times 10^{-8}\ TeV^{-2}$, the NP contribution is about an order of magnitude smaller than the present experimental bound at 2$\sigma$.

\subsection{Z pole constrains.}
\label{Z-pole}
For theoretical calculations of contribution from new fermions to the $Z$ decay into two fermions via higher order loops, we have used \cite{our2-ref23}
\begin{align}
Br(Z \rightarrow f_{i}f_{i}) = \frac{G_{F}}{3\sqrt{2}\pi}\frac{m_{Z}^{3}}{(16\pi^{2})^{2}\Gamma^{tot.}_{Z}}(T^{j}_{3} - Q^{j}\sin^{2}(\theta_{W}))^{2}|h_{i}^{'}|^{4}|[F_{2}(m_{k},m_{l}) + F_{3}(m_{k},m_{l})]|^{2}
\end{align}
where
\begin{align}
F_{2}(a, b) = \int^{1}_{0}dx(1-x)\ln{[(1-x)a^{2} + xb^{2}]}
\end{align}
and
\begin{align}
F_{3}(a, b) =& \int^{1}_{0}dx\int^{1-x}_{0}dy\frac{(xy-1)m_{Z}^{2} + (a^{2}-b^{2})(1-x-y) - \Delta\ln\Delta}{\Delta} \\
\Delta =& -xym^{2}_{Z} + (x+y)(a^{2}-b^{2}) + b^{2}
\end{align}
with $\Gamma^{tot}_{Z} = 2.4952$; $k = E\ or\ N$ and $l = \phi_{LQ}\ or\ \eta_{LQ}\ or\ \eta\ or\ S$ depending on the final state and Yukawa coupling involved and index $j$ refers to the exotic fermions in the loop.
At $m_{N} \approx m_{E_{h}} \approx m_{E_{l}} = 300$ GeV, $m_{\eta_{LQ}} = m_{\phi_{LQ}} = 900$ GeV, $m_{H^{0}} = 150$ GeV, $m_{A^{0}} = 200$ GeV and $m_{S} = 150$ GeV we have $Br(Z \rightarrow \bar{d}d) \approx 0$ due to Eqs(\ref{hd}) and $Br(Z \rightarrow \bar{u}u)^{NP}$, $Br(Z \rightarrow \bar{c}c)^{NP}$ $<<$ $Br(Z \rightarrow \bar{s}s)^{NP} \approx \mathcal{O}(10^{-10})$ compared to $Br(Z \rightarrow \bar{u}u)^{Exp} \approx Br(Z \rightarrow \bar{s}s)^{Exp} > Br(Z \rightarrow \bar{c}c)^{Exp} \approx 0.0021$, the NP contributions are negligible. With $h^{'}_{2b} \approx 3.52$ and $h^{'}_{1b} = 0.156 +1.290\times 10^{-4}i$ gives $Br(Z \rightarrow \bar{b}b)^{NP} \approx 3.15\times 10^{-5}$ which is an order of magnitude smaller than the experimental error where $Br(Z \rightarrow \bar{b}b)^{Exp}_{error} = 5\times 10^{-4}$ \cite{our2-ref1}. We take $h_{1e}$, $h_{2e}$, $h_{1\tau}$, $h_{2\mu}$ $<<$ 1 and $h_{1\mu} = h_{2\tau} = 3.52 \approx 2\sqrt{\pi}$. Then we get $Br(Z \rightarrow \bar{e}e)^{NP}$ $<<$ $Br(Z \rightarrow \bar{\mu}\mu)^{NP} = 2.369\times 10^{-6}$, $Br(Z \rightarrow \bar{\tau}\tau)^{NP} = 6.416\times 10^{-6}$ and $Br(Z \rightarrow \bar{\nu_{\tau}}\nu_{\tau})^{NP} = 2.080\times 10^{-5}$, compared to the respective experimental errors \cite{our2-ref1}, we have $Br(Z \rightarrow \bar{\mu}\mu)^{Exp}_{error} \approx 6.6\times 10^{-5}$, $Br(Z \rightarrow \bar{\tau}\tau)^{Exp}_{error} \approx 8.3\times 10^{-5}$ and $Br(Z \rightarrow \bar{\nu}\nu)^{Exp}_{error} \approx 5.5\times 10^{-4}$ which are an order of magnitude larger than the respective NP contributions and in all other cases the NP contributions are smaller than the respective experimental error estimates by two orders of magnitude or smaller and so negligible. Also there is a contribution to the muon (g-2) from the Yukawa coupling involving the two neutral components of the inert-doublet($\eta$) and give $\delta a_{\mu} = 1.152\times 10^{-9}$ \cite{our2}, which is within 2.1$\sigma$ of the experimental value whereas SM shows a deviation of about 3.6$\sigma$ from the experimental value \cite{our2-ref1}.

\subsection{Implications to $R(D^{(*)})$, $R_{K^{(*)}}$, neutrino masses, Baryon-genesis and DM.}
\label{subsect:bctnu}

As indicated in a recent model independent analysis of $R(D^{(*)})$ data with new estimates of the form factors \cite{ZRHuang}, the vector type NP is the best fit to the data while tensor type NP is highly restricted and scalar type NP is almost ruled out. In our model the terms in the Eqs(\ref{Yuk-Eq}) involving $E_{1}$ can not contribute to $b \rightarrow c \tau \nu_{\tau}$ but relevant terms involving $L_{2R}$ can contribute to the decay at box loop level given as \cite{our3}
\be
\mathcal{H}^{eff} = \frac{4G_{F}}{\sqrt{2}}V_{cb}(1 + C^{NP})[(c,b)(\tau ,\nu_{\tau})]
\ee
where $(c,b)(\tau ,\nu_{\tau})$ is the SM left handed vector four current operator and $C^{NP}$ is given as
\be
C^{NP} = \mathcal{N}\frac{(-V_{ub}h_{2d} - V_{cb}h_{2s} + V_{tb}h_{2b})|h_{2s}||h_{2\tau}|^{2}}{64\pi^{2}m_{E}^{2}}S(x_{i},x_{j}),
\ee
where $S(x_{i},x_{j}) = \frac{1}{(1-x_{i})(1-x_{j})} + \frac{x_{i}^{2}\ln{x_{i}}}{(1-x_{i})^{2}(x_{i}-x_{j})} - \frac{x_{j}^{2}\ln{x_{j}}}{(1-x_{j})^{2}(x_{i}-x_{j})}$ is the Inami-Lim functions \cite{our3}\cite{our3-ref11}\cite{our3-ref15} with $\frac{1}{\mathcal{N}} = \frac{4G_{F}|V_{cb}|}{\sqrt{2}}$, $x_{i} = \frac{m_{\phi_{LQ}}^{2}}{m_{E}^{2}}$, $x_{j} = \frac{m_{S}^{2}}{m_{E}^{2}}$ and $m_{E} \approx m_{N}$. Using the CKM matrix elements from the PDG \cite{our2-ref1} as in \cite{our3} and the benchmark values of the masses of the new particles and Yukawa couplings as before we have
\be
R(D)^{NP} = 0.31 \pm 0.167
\ee
and
\be
R(D^{*})^{NP} = 0.260 \pm 0.054,
\ee
which is within 1$\sigma$ of the theoretical and experimental errors combined \footnote{where theoretical errors are estimated by scaling the experimental errors by $\sqrt{\chi^{2}} = \sqrt{(} \frac{(R(D)^{Exp} - R(D)^{NP})^{2}}{\sigma^{2}_{Exp}(D)} + \frac{(R(D^{*})^{Exp} - R(D^{*})^{NP})^{2}}{\sigma^{2}_{Exp}(D^{*})})$}.
\begin{figure}[h!]
\begin{minipage}[t]{0.48\textwidth}
\hspace{0.4cm}
\includegraphics[width=2\linewidth, height=8cm]{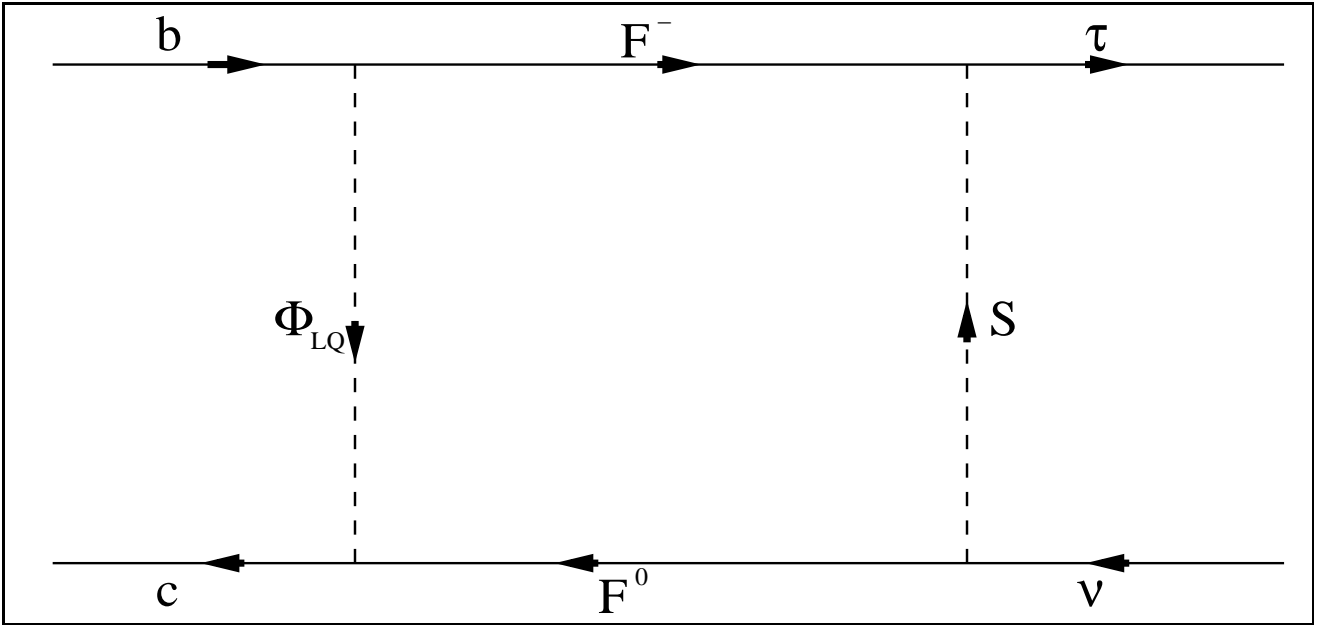}
\end{minipage}
\caption{Contributions to the $b \rightarrow c \tau \nu_{\tau}$ decay from the new particles at box loop level, where $F$ denoting the relevant new fermions involved.}
\label{Fig1:fig2}
\end{figure}
As mentioned above the Yukawa terms involving $L_{2R}$ can not contribute substantially to $B \rightarrow K^{(*)}\mu^{+}\mu^{-}$ but Yukawa terms involving $E_{1R}$ in Eqs(\ref{Yuk-Eq}) can contribute to this decay mode via box loop. In our model due to presence of terms involving $E_{1R}$, the NP contributions to Wilson coefficients $C_{9}^{NP}$ and $C^{NP}_{10}$ via box loop can be expressed as \cite{our2}\cite{our2-ref11}
\be
C^{NP}_{9} = -C_{10}^{NP} = N\frac{Re(h^{'}_{1b}h^{*'}_{1s})|h_{1\mu}|^{2}}{2\times 32\pi \alpha_{EM}m^{2}_{E}}[F(x_{\eta_{LQ}},x_{{H^{0}}}) + F(x_{\eta_{LQ}},x_{{A^{0}}})]
\ee
where $N^{-} = \frac{4G_{F}}{\sqrt{2}}V_{tb}V_{ts}^{*}$ and $F(x,y) = \frac{1}{(1-x)(1-y)} + \frac{x^{2}\ln{x}}{(1-x)^{2}(x-y)} + \frac{y^{2}\ln{y}}{(1-y)^{2}(y-x)}$ with $x_{\eta_{LQ}} = \frac{m_{\eta_{LQ}}^{2}}{m_{E}^{2}}$, $x_{H^{0}} = \frac{m_{H^{0}}^{2}}{m_{E}^{2}}$ and $x_{A^{0}} = \frac{m_{A^{0}}^{2}}{m_{E}^{2}}$ and as shown in section \ref{sub:Yukawa} $m_{E_{h}} \approx m_{E_{l}} = m_{E}$ is taken. Then with benchmark values of the new particle masses and Yukawa couplings in the flavor states implies $h^{'}_{1b}h^{'*}_{1s} = -0.027 + \mathcal{O}(10^{-5})i$ and gives
\be
C_{9}^{NP} = -C_{10}^{NP} = -0.46,
\ee
which is within 1.1$\sigma$ of the combine global best fit estimate of these NP Wilson coefficients to the data \cite{our2-ref9}. Besides the NP contributions to $B^{0}_{s}-\bar{B}^{0}_{s}$, see section \ref{subsect:Osc} for details, $C_{10}^{NP}$ can contribute to $Br(B_{S}^{0} \rightarrow \mu^{+}\mu^{-})$ which is measured to be consistent with the SM prediction of $Br(B_{S}^{0} \rightarrow \mu^{+}\mu^{-})^{SM} = (3.66 \pm 0.23)\times10^{-9}$ \cite{our2-ref28} compared to $Br(B_{S}^{0} \rightarrow \mu^{+}\mu^{-})^{Exp} = (3.28^{+0.7}_{-0.6})\times10^{-9}$ with $C^{SM}_{10} \approx -4.31$ and $C^{NP}_{10} \approx +0.46$, the $Br(B_{S}^{0} \rightarrow \mu^{+}\mu^{-})^{NP}$ is well within 1$\sigma$ of the experimental value, see \cite{our2-ref11}\cite{our2} for detail calculations. The NP contribution to $(C_{7} + 0.24C_{8})^{NP} \approx \mathcal{O}(10^{-3})$ which affect the $b \rightarrow s\gamma$ rate and the NP contribution is about 2 orders of magnitude smaller than the present experimental bound at 2$\sigma$ \cite{our2-ref11}, also check \cite{our2} for detail calculations. The bound coming from $Br(B \rightarrow K^{(*)}\bar{\nu}\nu)^{Exp}$ on NP is much weaker than that from $Br(B \rightarrow K^{(*)}\bar{\mu}\mu)^{Exp}$ and so constrains from these modes are automatically satisfied \cite{our2-ref11}.\\
Similar to the estimates in \cite{our3}, the NP contributions to $B_{c} \rightarrow \tau \nu_{\tau}$, $D_{s} \rightarrow \tau \nu_{\tau}$, $\tau^{\pm} \rightarrow K_{s}^{0}\pi^{\pm}$ including to CP violations due to $h_{2d}$ being complex, $b \rightarrow s\gamma$, $B \rightarrow K^{(*)}\tau^{+}\tau^{-}$, $B_{s} \rightarrow \tau^{+}\tau^{-}$ and $D^{0} \rightarrow (\pi^{0})\bar{\nu}_{\tau}\nu_{\tau}$ are all negligible compare to the respective experimental bounds \cite{our2-ref1}. The NP contributions to the anomalous magnetic moment of $\tau$ is $\delta a_{\tau} \approx \mathcal{O}(10^{-8})$ which is many orders of magnitude smaller than the latest experimental bound of $-0.052 < \delta a_{\tau} < 0.013$\cite{our2-ref1}.\\
Now with the introduction of heavy righthanded neutrinos $N_{(e,\mu,\tau)R}$ there can be Yukawa terms such as
\be
\mathcal{L}_{\nu} = \sum^{\tau}_{i,j=e}h_{ij}\bar{L}_{i}i\sigma_{2}\eta N_{jR} + h.c,
\ee
which can give Majorana mass term of $M_{\alpha\beta}\bar{\nu}_{\alpha}\nu_{\beta}$ for the light neutrinos with \cite{our2-ref15}
\be
M_{\alpha\beta} = \sum_{i}\frac{h_{\alpha i}h_{\beta i}M_{iR}}{16\pi^2}[\frac{m_{H^{0}}^{2}}{m_{H^{0}}^{2}-M_{iR}^{2}}\ln{\frac{m_{H^{0}}^{2}}{M^{2}_{iR}}} - \frac{m_{A^{0}}^{2}}{m_{A^{0}}^{2}-M_{iR}^{2}}\ln{\frac{m_{A^{0}}^{2}}{M^{2}_{iR}}}]
\ee
where $M_{iR}$ being the masses of the heavy Majorana neutrinos. Then with benchmark values of masses $m_{H^{0}} = 150$ GeV, $m_{A^{0}} = 200$ GeV and taking lightest of $M_{iR} \geq 2.6\times 10^7$ GeV, we get $M_{\alpha\beta} = \mathcal{O}(0.01)$ eV for $|h_{\alpha i}h_{\beta i}| \approx 10^{-7}$. At this values of the parameters, generation of universe's Baryon access via Leptogenesis is also possible, see \cite{our2-ref15}\cite{our2-ref31} for more details. Although $H^{0}$ being one of the LSP, due to its large Yukawa couplings required from $R_{K^{(*)}}$ and $(g-2)_{\mu}$ data and non-observation of stable heavy charged particle in colliders etc. its contribution to the DM relic density would be small \cite{our2-ref23}\cite{our2-ref32}. Another trivial extension of our model is to include a new singlet scalar DM, although the Higgs portal of this DM is ruled out due to over abundance problem \cite{our2-ref24}\cite{our2-ref25}\cite{our2-ref26}\cite{our2-ref27}\cite{our2-ref16}, in our model there are many more new particles it can couple to so as to generate enough DM annihilation to avoid over abundance problem unlike SM Higgs only portal, see e.g \cite{our2} for an exotic scalar portal extension of the scalar singlet DM which can be easily incorporated into our model.

\section{Conclusions.}
\label{sect:conclusions}

In this work we have proposed an extension of SM lepton content by introducing a right handed and a left handed pair of $SU(2)_{L}$ doublet leptons ($L_{1L},L_{2R}$) along with their respective charged right handed and left handed $SU(2)_{L}$ singlet partners ($E_{1R},E_{2L}$), plus we also added three heavy right handed neutrinos ($N_{eR},N_{\mu R},N_{\tau R}$) to generate small neutrino masses at loop level. We extended the SM Higgs sector by introducing two $SU(2)_{L}$ doublet leptoquarks ($\phi_{LQ}, \eta_{LQ}$) along with an inert-Higgs-doublet ($\eta$) and a complex singlet scalar S, plus a real singlet scalar ($\phi$) whose VEV gives the dominant masses to the new leptons. All the new particles are assumed to be odd under a $\mathcal{Z}_{2}$ except $\phi$ which is assumed to be even so that it can develop a non-zero VEV. With these new particles added to the SM, we have shown that all the observed anomalies in lepton universality observables in semi-leptonic B meson decays can be explained with satisfying constrains from nuetral meson oscillations, precision Z-pole data, etc. within reasonable error limits. In addition our model is also able to explain the small neutrino masses along with generations of universe's Baryon excess via leptogenesis. Also our model have enough new parameters to avoid over abundance problem in scalar singlet DM and so DM can be incorporated trivially in our model.


{\large Acknowledgments: \large} This work was partially supported by funding available from the Department of Atomic Energy, Government of India, for the Regional Centre for Accelerator-based Particle Physics (RECAPP), Harish-Chandra Research Institute.

\end{document}